\documentclass[twocolumn]{article}

\usepackage[super,sort&compress,comma]{natbib}
\usepackage{graphicx} 
\usepackage{dcolumn}
\usepackage{bm}
\usepackage{amsmath}
\usepackage{amssymb}
\usepackage[none]{hyphenat}
\usepackage{url}
\usepackage[left=1.5cm, right=1.5cm, top=1.785cm, bottom=2.0cm]{geometry}

\makeatletter
\renewcommand\LARGE{\@setfontsize\LARGE{15pt}{17}}
\renewcommand\Large{\@setfontsize\Large{12pt}{14}}
\renewcommand\large{\@setfontsize\large{10pt}{12}}
\renewcommand\footnotesize{\@setfontsize\footnotesize{7pt}{10}}
\makeatother

\begin{document}

\twocolumn[
  \begin{@twocolumnfalse}

\begin{tabular}{m{4.5cm} p{13.5cm} }

 & \noindent\LARGE{\textbf{Interplay between magnetic order and electronic band structure in ultrathin GdGe$_2$ metalloxene films$^\dag$}} \\
\vspace{0.3cm} & \vspace{0.3cm} \\

 & \noindent\large{
 Andrey V. Matetskiy,$^{\ast}$\textit{$^{a,b}$}
 Valeria Milotti,$^{\star}$\textit{$^{a}$}
 Polina M. Sheverdyaeva,$^{\diamond}$\textit{$^{a}$}
 Paolo Moras,\textit{$^{a}$}
 Carlo Carbone,\textit{$^{a}$}
 and Alexey N. Mihalyuk$^{\bullet}$\textit{$^{b,c}$}} \\

\includegraphics{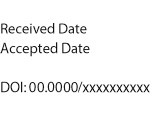} & \noindent\normalsize{
Dimensionality can strongly influence the magnetic structure of solid systems. Here, we predict theoretically and confirm experimentally that the antiferromagnetic (AFM) ground state of bulk gadolinium germanide metalloxene, which has a quasi-layered defective GdGe$_2$ structure, is preserved in the ultrathin film limit. \textit{Ab initio} calculations demonstrate that ultrathin GdGe$_2$ films present in-plane intra-layer ferromagnetic coupling and AFM inter-layer coupling in the ground state. Angle-resolved photoemission spectroscopy finds the AFM-induced band splitting expected for the 2 and 3 GdGe$_2$ trilayer (TL) films, which disappear above the Néel temperature. The comparative analysis of isostructural ultrathin DyGe$_2$ and GdSi$_2$ films confirms the magnetic origin of the observed band splitting. These findings are in contrast with the recent report of ferromagnetism in ultrathin metalloxene films, which we ascribe to the presence of uncompensated magnetic moments.
} \\

\end{tabular}

 \end{@twocolumnfalse} \vspace{0.6cm}

  ]

\renewcommand*\rmdefault{bch}\normalfont\upshape
\rmfamily
\section*{}
\vspace{-1cm}

\footnotetext{\textit{$^{a}$ Istituto di Struttura della Materia-CNR (ISM-CNR), Strada Statale 14 km 163.5, 34149, Trieste, Italy}}
\footnotetext{\textit{$^{b}$ Institute of Automation and Control Processes FEB RAS, 690041 Vladivostok, Russia}}
\footnotetext{\textit{$^{c}$ Institute of High Technologies and Advanced Materials, Far Eastern Federal University, 690950 Vladivostok, Russia}}
\footnotetext{$^{\ast}$E-mail: andrei.matetskii@trieste.ism.cnr.it}
\footnotetext{$^{\star}$E-mail: valeria.milotti@unipd.it. Present affiliation: Department of Physics and Astronomy "Galileo Galilei", University of Padua, Via F. Marzolo, 8, 35131 Padua, Italy}
\footnotetext{$^{\diamond}$E-mail:  polina.sheverdyaeva@ism.cnr.it}
\footnotetext{$^{\bullet}$E-mail: mih-alexey@yandex.ru}

\footnotetext{\dag~Electronic Supplementary Information (ESI) available: [Interplay between magnetic order and electronic band structure in ultrathin GdGe$_2$ metalloxene films]. See DOI: 00.0000/00000000.}

\section{Introduction}
The scientific community working in the field of magnetism shows growing interest towards interfaces, surfaces and materials of reduced dimensionality \cite{Vedmedenko2020}. As an example, the recently discovered two dimensional (2D) magnetic materials are now the focus of intensive research efforts \cite{Jiang2021}. In these layered compounds, including Fe$_3$GeTe$_2$\cite{Burch2018, Fei2018}, Cr$_2$Ge$_2$Te$_6$ \cite{Gong2017}, CrI$_3$ \cite{McGuire2015, Huang2017}, the ferromagnetic (FM) state is stabilized by the intrinsic anisotropy of the crystal structure that reduces the spin degree of freedom and allows to overcome the restriction of the Mermin-Wagner theorem \cite{Gibertini2019}. They are an ideal experimental test environment for the verification of 2D magnetic phase-transition theories \cite{Peierls1936, Mermin1966, Hohenberg1967}. \par
Another important research topic connected to the previous one is  non-collinear magnetism, which gives rise to skyrmions, for instance \cite{Zhang2022, Liu2020, Ying2022, Deng2020, Ge2020}. Among the various compounds showing non-collinear magnetism, silicides and germanides of the $3d$ elements with the cubic B20 structure attracted remarkable attention \cite{Tokura2021}. The observation of extremely small skyrmions in the Gd-based inter-metallic compounds has also been reported \cite{Kurumaji2019, Khanh2020}. \par
The binary silicides and germanides of the rare-earth elements have been the subject of intense study since 1960s \cite{RE_book}. These compounds have a quasi-layered defective AlB$_2$ crystallographic structure in which honeycomb-like semiconductor layers are separated by rare-earth atomic layers (trilayer, TL). By analogy with materials containing honeycomb layers they are called metalloxenes. They can be epitaxially grown on the parent semiconductor substrates \cite{ENGELHARDT2006755} to form hetero-structures with low Schottky-barrier heights \cite{Vandre1999}. Nowadays, due to growing interest in 2D and graphene-like structures, these systems are being revisited \cite{Wanke2009SS, Franz2016, Parfenov2019, Tokmachev2019, Averyanov2020}. \par
In the bulk most of the metalloxenes tend to be antiferromagnetic (AFM) with N\'{e}el temperatures (T$_\text{N}$) in the range of 10$\div$50 K \cite{RE_book}. For example, T$_\text{N}$ =38 K and 54 K for Gd germanides \cite{Tokmachev2019MH}  and silicides \cite{Roger2006}, respectively. Surprisingly, 2D  FM order with magnetic moments in the order of 0.1 \textmu$_\text{B}$  has been recently found in ultrathin metalloxenes films \cite{Tokmachev2019, Averyanov2020}.  The 2D electron confinement could explain the experimental observation of the AFM to FM transition with decreasing the film thickness. However, the origin of this behavior has not been fully clarified and deserves further analysis. \par
In this regard, temperature-dependent angle-resolved photoemission spectroscopy (ARPES) is a powerful tool, since it remains sensitive to the  deviations of band structure induced by magnetic order also in low-dimensional systems \cite{Pakhira2023, Chikina2014, Schmitt2019, Elmers2023}. In particular, ARPES, in combination with \textit{ab initio} density-functional calculations (DFT), can give a clue on the exact magnetic structure \cite{Lee2023, eremeev2023insight, Fernandez2020}. In the present study we investigate by ARPES and DFT the electronic and magnetic structure of Gd germanide metalloxene films grown on Ge(111). Our results show that the AFM order remains the ground state of the films down to the ultrathin limit of 2 and 3 TL. The FM order observed in Ref. \cite{Tokmachev2019MH}  turns out to be caused by the presence of uncompensated magnetic moments in the order of 0.1 \textmu$_\text{B}$. These moments are found to derive from the hybridization of Gd $5d$ and Ge $4p$ levels in the low-symmetry film structure and/or are ascribed to the coexistence of multiple film thickness. ARPES data of Gd silicide and Dy germanide ultrathin metalloxene films are provided to strengthen our conclusions.\par

\section{Experimental and calculation details}
The experiments were performed at the VUV-Photoemission beamline at Elettra synchrotron (Trieste, Italy), by means of ARPES and low-energy electron diffraction (LEED) methods. The base pressure of the analysis and preparation chambers was $\leq$ 1.0$\times$10$^{-10}$ Torr and $\leq$ and 3$\times$10$^{-10}$ Torr, respectively.  The Ge(111) and Si(111) substrates  were used for metalloxene films growth. The Ge(111) substrates were sputtered with Ar ion bombardment and then annealed at 650$^{\circ}$C; this procedure was repeated several times until the sharp c(2$\times$8) LEED pattern was obtained. In order to obtain Si(111)7$\times$7 surface reconstruction, Si(111) was flash annealed to a temperature of $\sim$1200$^{\circ}$C. Gd  and Dy were deposited using electron bombardment sources with rates of  $\sim$0.25 ML/min [1 monolayer (ML) = 6.2$\times$10$^{14}$ cm$^{-2}$ in terms of the Ge(111) surface atomic density]. The evaporation rate was calibrated by the observation of LEED patterns that correspond to known surface reconstructions: 5$\times$2 at coverage < 1 ML, 1$\times$1 that correspond to completion of the first TL at coverage $\sim$1 ML and  $\sqrt{3}\times\sqrt{3}$ at coverage above 1 ML \cite{ENGELHARDT2006755}. During deposition the substrates were held at $\sim$400$^{\circ}$C. The Si substrates were annealed at $\sim$550-650$^{\circ}$C after Gd deposition in order to the improve the crystalline order of the films. It should be pointed out here that such procedure gives rise to films with multiple film thickness after the completion of 1 TL \cite{ENGELHARDT2006755, Wanke2009SS}, as it will be shown while discussing the properties of the 2 TL films. ARPES measurements were conducted in the 14$\div$82 K temperature range using a  Scienta R4000 electron analyzer and excitation energies between 25\,eV and 55\,eV with linearly polarized light. The electron spectrometer was placed at 45$^{\circ}$ with reference to the direction of the incoming photon beam. The labels of the high symmetry points in the ARPES spectra refer to the 1$\times$1 surface Brillouin zone (BZ) of the substrates. \par
Calculations were based on DFT as implemented in the Vienna \textit{ab initio} simulation  package VASP.\cite{VASP1} The projector-augmented wave approach\cite{PAW} was used to describe the electron-ion interaction and the generalized gradient approximation (GGA) of Perdew, Burke, and Ernzerhof (PBE)\cite{PBE} was employed as the exchange-correlation functional. The scalar relativistic effect and the spin-orbit coupling (SOC) were taken into account. To simulate the GdGe$_{2}$, DyGe$_{2}$ and GdSi$_{2}$ structures we used a slab consisting of four bilayers (BL) of germanium/silicon with the PBE-optimized bulk lattice constants. Hydrogen atoms were used to passivate the dangling bonds at the bottom of the slab. The atomic positions of adsorbed Gd/Dy atoms and atoms of upper Ge/Si layer and layers within the three BLs of the slab were optimized. Substrate atoms of the deeper layers were kept fixed at the bulk crystalline positions. The kinetic cutoff energy was 250 eV, and a 12$\times$12$\times$1 and 7$\times$7$\times$1 $k$-point meshes were used to sample the 1$\times$1 and $\sqrt3\times\sqrt3$ supercell BZ, respectively. The geometry optimization was performed until the residual force on atoms was smaller than 10 meV/\AA.
For band-structure calculations, two types of Gd/Dy pseudopotentials were used \cite{VASP2}. The trivalent Gd/Dy potentials, where strongly localized, valence 4$f$ electrons are treated as core states were used for non-magnetic band-structure calculations. In order to describe the magnetic properties, the standard Gd/Dy potentials were used for spin-polarized non-collinear calculations, in which the $f$ electrons are treated as valence states. The Heyd-Scuseria-Ernzerhof (HSE06) screened hybrid functional was used to accurately calculate the Ge gap and to avoid the self-interaction errors arising from an incorrect description of partially filled $f$ states of Gd/Dy \cite{HSE}. \par 

\section{Results and Discussion}
\subsection{Atomic structure of the ultrathin GdGe$_2$ films}
\begin{figure}[t]
	\centering
	\includegraphics[width=1.0\columnwidth]{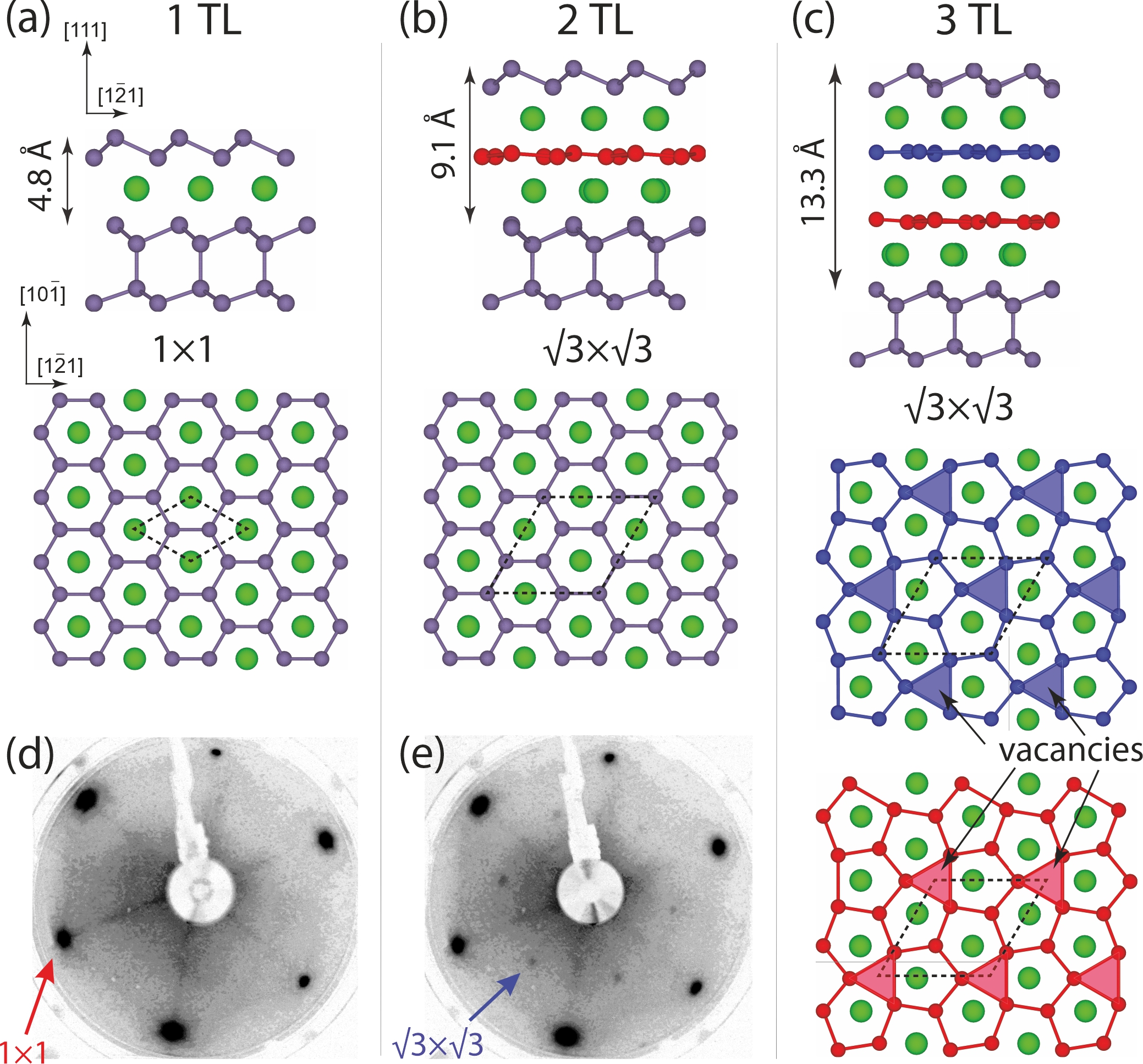}
	\caption{(a-c) Side and top views of the relaxed  atomic structure models of the GdGe$_2$ films with thickness of 1-3 TL placed on Ge(111) slab. Green balls correspond to Gd atoms while gray, red and blue balls correspond to Ge. Black dashed rhombuses outline the 1$\times$1 and $\sqrt{3}\times\sqrt{3}$ unit cells. (d,e) LEED patterns measured for 1 and 2 TL GdGe$_2$ films, respectively.	}
    \label{f:structure}
\end{figure}

Trivalent rare-earth elements form germanides and silicides of various stoichiometries \cite{RE_book}. In the present paper we will study defective AlB2$_2$  type ultrathin Gd germanide metalloxene films. Figure \ref{f:structure}(a-c) shows the relaxed crystallographic structures of 1-3 TL films on Ge(111) obtained by the \textit{ab initio} random structure searching (AIRSS) method \cite{Pickard2011}. The atomic structure of the 1 TL GdGe$_2$ (Fig.~\ref{f:structure}(a)) corresponds to the one reported previously in the literature \cite{Sanna2021M,Wanke2009SS,Tokmachev2019MH}. It consists of a single layer of Gd atoms sandwiched between the Ge(111) substrate and the buckled Ge surface bilayer (BL). This BL displays a reversed buckling with respect to the substrate. The 1 TL system has GdGe$_2$ stoichiometry, hexagonal symmetry and 1$\times$1 LEED pattern (Fig.~\ref{f:structure}(d)). \par
The formation of films thicker than 1 TL reduces the symmetry of the surface from hexagonal to trigonal and changes the surface periodicity from 1$\times$1 to $\sqrt{3}\times\sqrt{3}$ \cite{Wanke2009SS}, as demonstrated by the emergence of weak $\sqrt{3}\times\sqrt{3}$ reflexes in the LEED pattern (Fig.~\ref{f:structure}(e)). This weak periodicity was connected with the formation of the vacancy lattice in the inner semiconductor BLs of metalloxenes \cite{Wetzel1996}. Due to these vacancies the Ge BLs become almost flat (red and blue balls in Figs.~\ref{f:structure}(b, c)) and the overall stoichiometry of the inner TLs becomes Gd$_3$Ge$_5$. In spite of this change, for convenience we will use the GdGe$_2$ notation for all Gd germanide films (we will adopt a similar nomenclature for Gd silicide and Dy germanide films), irrespectively of their thickness. \par
According to our calculations, the Gd atoms of the different layers are always located at the T$_{4}$ site with reference to the underlying Ge(111) substrate. This is in line with previous high-angle annular dark-field transmission electron microscopy observations \cite{Tokmachev2019MH,Parfenov2019}. The geometry of the topmost Ge BL is the following: the upper Ge atom is located in the T$_{1}$ site, while the lower Ge atom is located in the H$_{3}$ site. For the 2 TL system the most energetically stable atomic configuration of the intermediate flat Ge layer has a vacancy defect located in one of the T$_{1}$ sites. Importantly, the vacancy defects within flat Ge layers in 3 TL and thicker films have alternating positions with respect to the neighboring Ge layers, as shown in Fig.~\ref{f:structure}(c). The total thicknesses of 1, 2 and 3 TL GdGe$_2$ films is 4.8, 9.1, and 13.3 \AA, respectively.\par

\subsection{Thickness-dependent electronic band structure of the GdGe$_2$ films in the paramagnetic phase}
\begin{figure*}[ht!]
\centering
  \includegraphics[width=1.0\textwidth]{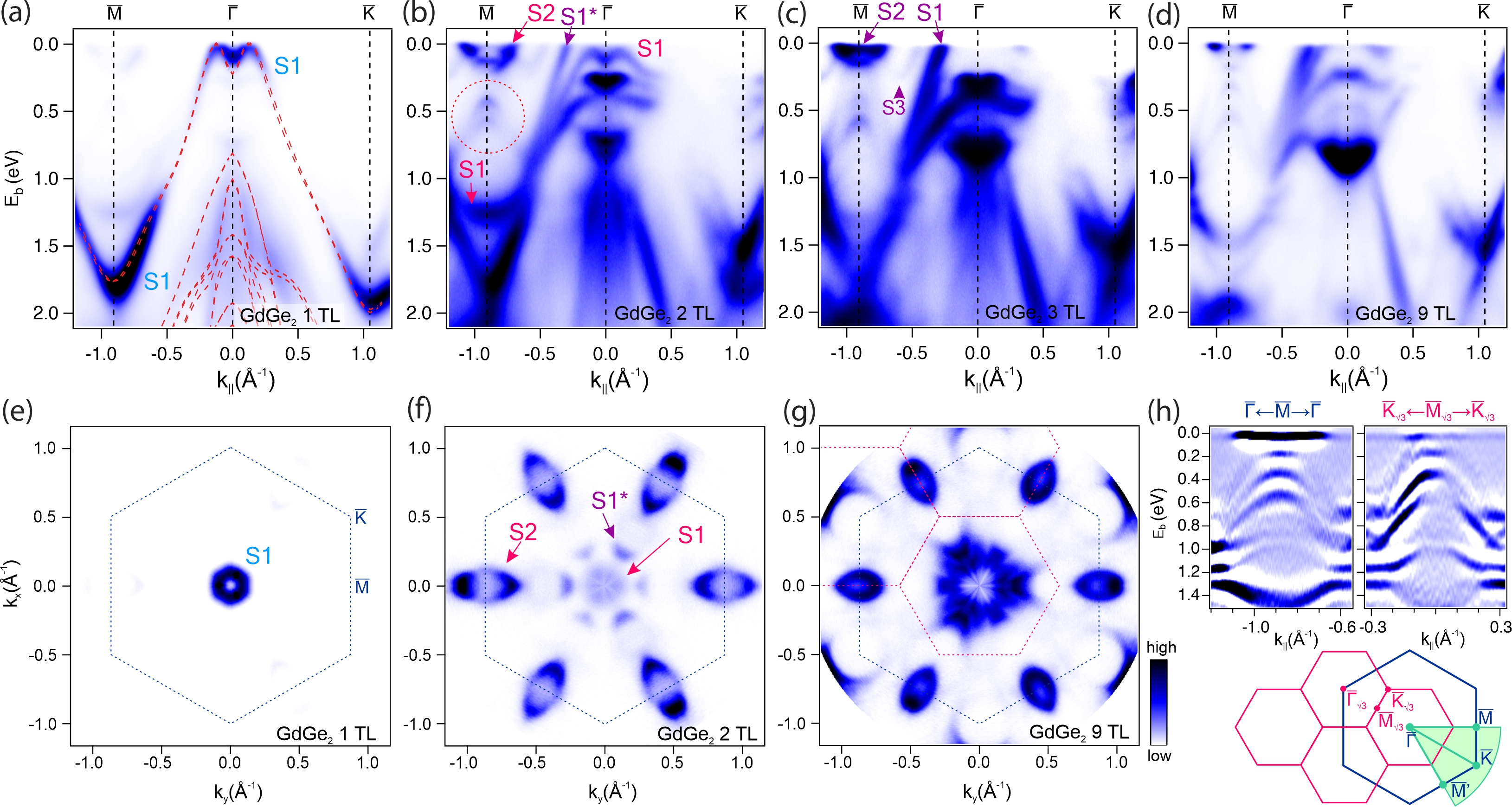}
\caption{(a-d) ARPES spectra of GdGe$_2$ films with thicknesses of 1, 2, 3 and 9 TL, respectively. All spectra are taken at 82 K with 35 eV photon energy. In (a) the results of DFT calculation for 1 TL GdGe$_2$ are overlaid on the ARPES data with adjusted Fermi level. (e-g)  Fermi surface maps of GdGe$_2$ films with thicknesses of 1, 2 and 9 TL, respectively.  The color of the labels S1, S2 and S3 indicates a specific film thickness: blue - 1 TL, red - 2 TL, violet - 3 TL. (h) Top: second derivatives of the ARPES spectra in the vicinity of $\bar{\text{M}}$ and $\bar{\text{M}}_{\sqrt{3}}$ points. Bottom: schematics of the surface BZs for the 1$\times$1 and $\sqrt{3}\times\sqrt{3}$ lattices and  ARPES measurement geometry. Green lines mark the direction for the experimental measurements in (a-d). The green sector depicts the azimuthal scan used to obtain the Fermi surfaces in (e-g). }
\label{f:thickness}
\end{figure*}

In this section we show ARPES data and DFT calculations for GdGe$_{2}$ films of different thickness. The ARPES spectra for 1, 2, 3 and 9 (nominal) TL are shown in Figs.~\ref{f:thickness}(a-d). Colored arrows and S1-S3 labels mark the characteristic features in the band structure that allow the identification of the film thickness and are discussed below in connection with the magnetic properties. The electronic structure of 1 TL (Fig.~\ref{f:thickness}(a)) consists in the highly dispersive S1 surface band. This band is reproduced well by the DFT calculations for the 1 TL GdGe$_2$ model previously discussed in section 3.1 (DFT results are overlaid onto the ARPES data in Fig.~\ref{f:thickness}(a)). The Gd-Ge hybridization leads to the formation of the camel-like feature at the top of the S1 band, which touches the Fermi level and forms a ring-like hole pocket around the $\bar{\Gamma}$ point (Fig.~\ref{f:thickness}(e)). \par
The 2 TL GdGe$_2$ film (Figs.\ref{f:thickness}(b)) shows a S1 hole-like band similar to the 1 TL case. With respect to 1 TL, it is shifted to slightly higher binding energies at $\bar{\Gamma}$ and by 0.5 eV towards lower energies  at the $\bar{\text{M}}$ point and is fully occupied. Additionally, a new S2 band leads to the formation of the electron pocket around the $\bar{\text{M}}$ points (Figs.\ref{f:thickness}(b,f)). Below this pocket a faint cone-like feature appears (marked by red circle in Figs.\ref{f:thickness}(b)). A metallic band labeled S1$^{\ast}$ is observed along the $\bar{\Gamma} -  \bar{\text{M}}$ direction. According to the calculation in Fig.S1 of the Electronic Supplementary Information (ESI) and Fig.\ref{f:triple}(e) it is associated to the 3 TL film and, therefore, is a manifestation of the coexistence of multiple thickness above the completion of 1 TL. Further confirmation of this assignment comes from the fact that at the nominal coverage of 3 TL the intensity of the S1 band, whose dispersion coincides with S1$^{\ast}$, is much stronger (Fig.\ref{f:thickness}(c)). Here, a weak electron pocket labeled S3 can be observed between $\bar{\Gamma}$ and $\bar{\text{M}}$.  \par
In thicker films (Fig.\ref{f:thickness}(d, g)) the spectrum in the vicinity of Fermi level is characterized by the presence of electron-like bands near the $\bar{\text{M}}$ point and hole-like bands near the $\bar{\Gamma}$ point. The apex of the cone-like band in $\bar{\text{M}}$ point shifts towards lower binding energies with thickness and start to hybridize with  the S2-like bands. This feature follows the vacancy-associated super-periodicity and can be found also in $\bar{\text{M}}_{\sqrt{3}}$ points (Fig.\ref{f:thickness}(h)). Therefore, while the hole pockets around $\bar{\Gamma}$ point and electron pockets in the $\bar{\text{M}}$ points remain dominant, the spectral intensities in thicker  films display additional features associated with the $\sqrt{3}\times\sqrt{3}$ lattice  which may be seen in the ARPES plot and Fermi surface map (Fig.\ref{f:thickness}(g)). \par
\begin{figure*}[ht!]
\centering
  \includegraphics[width=1.0\textwidth]{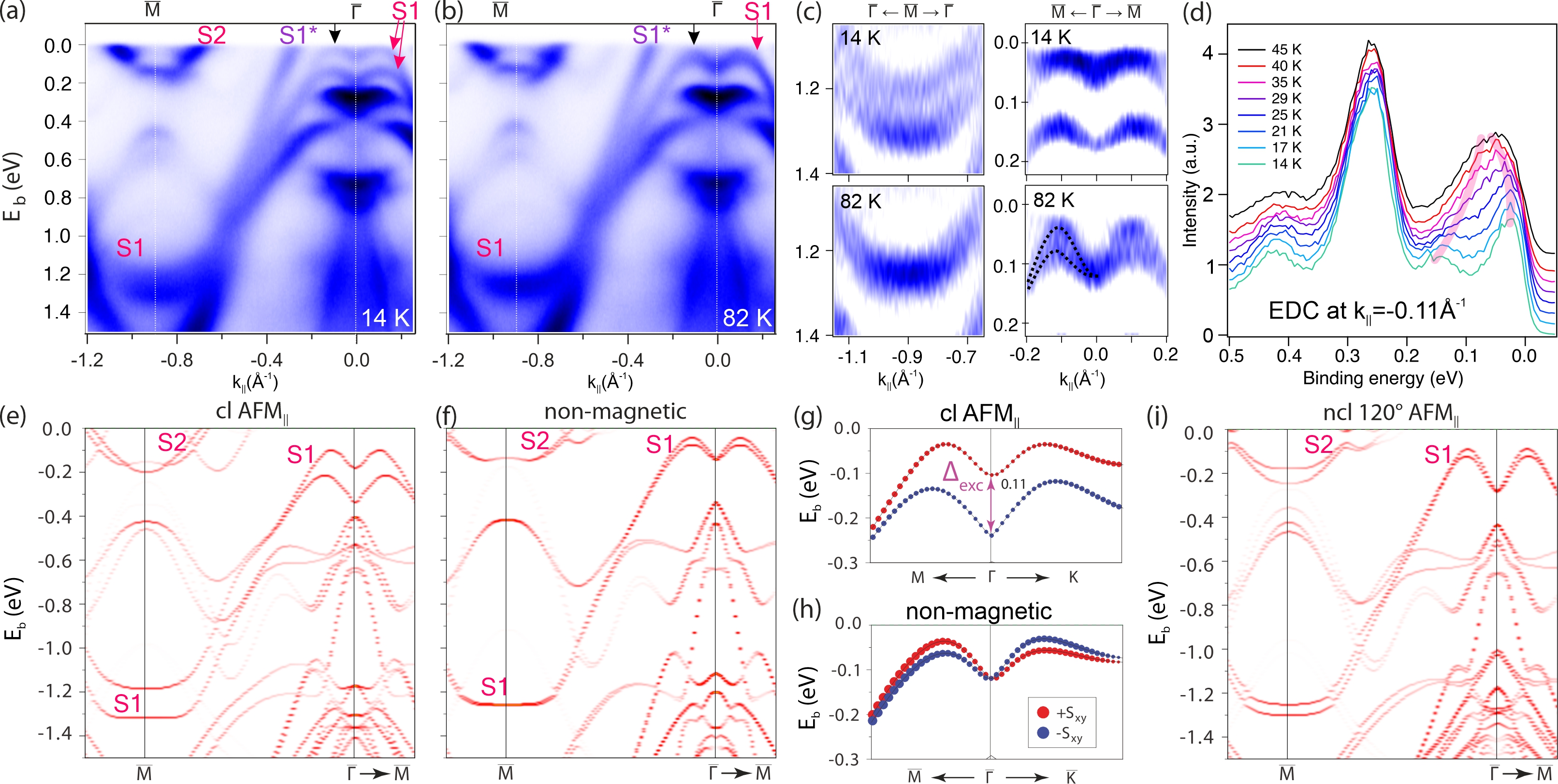}
\caption{ Electronic structure of the 2 TL  GdGe$_2$ film. (a, b) ARPES spectra taken at 14 K and 82 K, respectively. (c)  Second derivatives of the ARPES spectra in the vicinity of $\bar{\Gamma}$ and $\bar{\text{M}}$ points at 14 K (top row) and 82 K (bottom row). (d) Energy distribution curves taken at the top of the S1 band (k$_{||}$=0.11 \AA$^{-1}$, black arrows in (a, b)) in the temperature range 14$\div$45 K. The pink lines highlight the change of the two spin-split  bands with the temperature.  These bands do not overlap above T$_\text{N}$ due to the Rashba effect.  (e, f) DFT electronic structure calculations for 2 TL  GdGe$_2$ with cl AFM$_\|$ and non-magnetic configurations, respectively. The bands are unfolded onto the 1$\times$1 surface BZ to be easily compared with  the ARPES data. (g, h) Zoom of panels (e, f) near the top of the S1 band. Red and blue colors represent the opposite in-plane spin components. (i) DFT electronic structure calculations for 2 TL GdGe$_2$ with ncl 120${}^\circ$ AFM$_\|$ order unfolded onto the 1$\times$1 surface BZ.}
\label{f:double}
\end{figure*}
\subsection{Transition temperature and magnetic order of the GdGe$_2$ films}
In order to study the magnetic order and the effect of magnetism on the electronic structure of the GdGe$_2$ films, we performed comprehensive \textit{ab initio} calculations. The magnetic ground state of the Gd germanide system was determined by calculating the total energies for various magnetic configurations (see Tab.~\ref{tab1}). Collinear in-plane AFM (cl AFM$_\|$) represent the ground state for the multilayered films. The out-of-plane collinear AFM (cl AFM$\bot$) configuration is less favorable by 0.2 meV per Gd atom. While the difference in total energy between cl AFM$_\|$ and cl AFM$\bot$ is small it is enough to judge on magnetic ground state. The in-plane non-collinear 120${}^\circ$ AFM (ncl 120${}^\circ$ AFM$_\|$) has 5 meV higher energy than the ground state. Finally, the in-plane FM (FM$_\|$) and out-of-plane FM (FM$\bot$) configurations are much less favorable than cl AFM$_\|$. For 1 TL GdGe$_{2}$ film, FM$_\|$ is the ground state and is lower than FM$\bot$ by 0.5 meV.\par 
\begin{table}[htbp]
\begin{center}
\begin{tabular}{|c|c|c|c|c|c|}
\hline 
   cl AFM$_\|$&cl AFM$\bot$&ncl 120${}^\circ$ AFM$_\|$&FM$_\|$  & FM$\bot$\\
\hline
   0.0 &  0.2   & 5.0    & 10.0  &  11.0    \\
\hline
\end{tabular}
\end{center}
\caption{Energies (meV/Gd atom) of different magnetic configurations of the 2 TL GdGe$_2$ film with respect to the ground state. Energies for films of higher thicknesses follow similar trend.
[collinear in-plane AFM (cl-AFM$_\|$), collinear out-of-plane AFM (cl-AFM$\bot$); non-collinear 120${}^\circ$ AFM (ncl 120${}^\circ$ AFM$_\|$), in-plane FM (FM$_\|$), out-of-plane FM (FM$\bot$)]}
\label{tab1}
\end{table}
Let us now examine the effect of magnetism on the electronic band structure of the GdGe$_2$ films  by temperature-dependent ARPES and magnetic DFT calculations. We did not find any evidence of the magnetic behavior in the 1 TL GdGe$_2$ film down to  $\sim$14 K. We did not observe any magnetically-induced splitting of the S1 surface band or any other band structure changes, which were predicted by the  band structure calculations in the FM$_\|$ phase (Fig.S1, ESI). Therefore we come to the conclusion that the Curie temperature for the 1 TL lays below 14 K, in line with the magnetic measurements \cite{Tokmachev2018NC}.\par
We made a temperature-dependent ARPES measurements for the 2 TL GdGe$_2$ across the transition temperature reported in Ref.\cite{Tokmachev2019MH} and investigated the behavior of the S1 and S2 bands. Figures~\ref{f:double}(a, b) show the experimental spectra recorded at 14 and 82 K, respectively. At 14 K the S1 band displays a magnetic spin splittings of $\approx$130 meV at the $\bar{\Gamma}$ point just below the Fermi level and of $\approx$100 meV at the $\bar{\text{M}}$ point at E$_b\approx$1.25 eV. Both splittings  can be resolved in the second derivative spectra shown in  Fig.~\ref{f:double} (c). When the temperature is raised up to 82 K, the above mentioned spin splittings at $\bar{\Gamma}$  and $\bar{\text{M}}$ disappear, while a tiny k-dependent splitting can still be observed near the $\bar{\Gamma}$ point (Fig.\ref{f:double}(c)). Instead, the S2 state does not show any significant change with temperature. Thus, to track the magnetic transition of the 2 TL film we use the temperature dependence of the splitting of the S1 state near the $\bar{\Gamma}$ point. According to ARPES measurements made within the [14$\div$45] K temperature range the N{\'{e}}el transition temperature in 2 TL GdGe$_2$ film is 31$\pm$1 K (fitting results for the exchange splitting value in Fig.S2 deduced from energy distribution curves in Fig.~\ref{f:double}(d)). This indicates that already at the thickness of 2 TLs the GdGe$_2$ film  has a transition temperature that is close to the bulk value \cite{RE_book}. \par
\begin{figure*}[ht!]
\centering
  \includegraphics[width=0.8\textwidth]{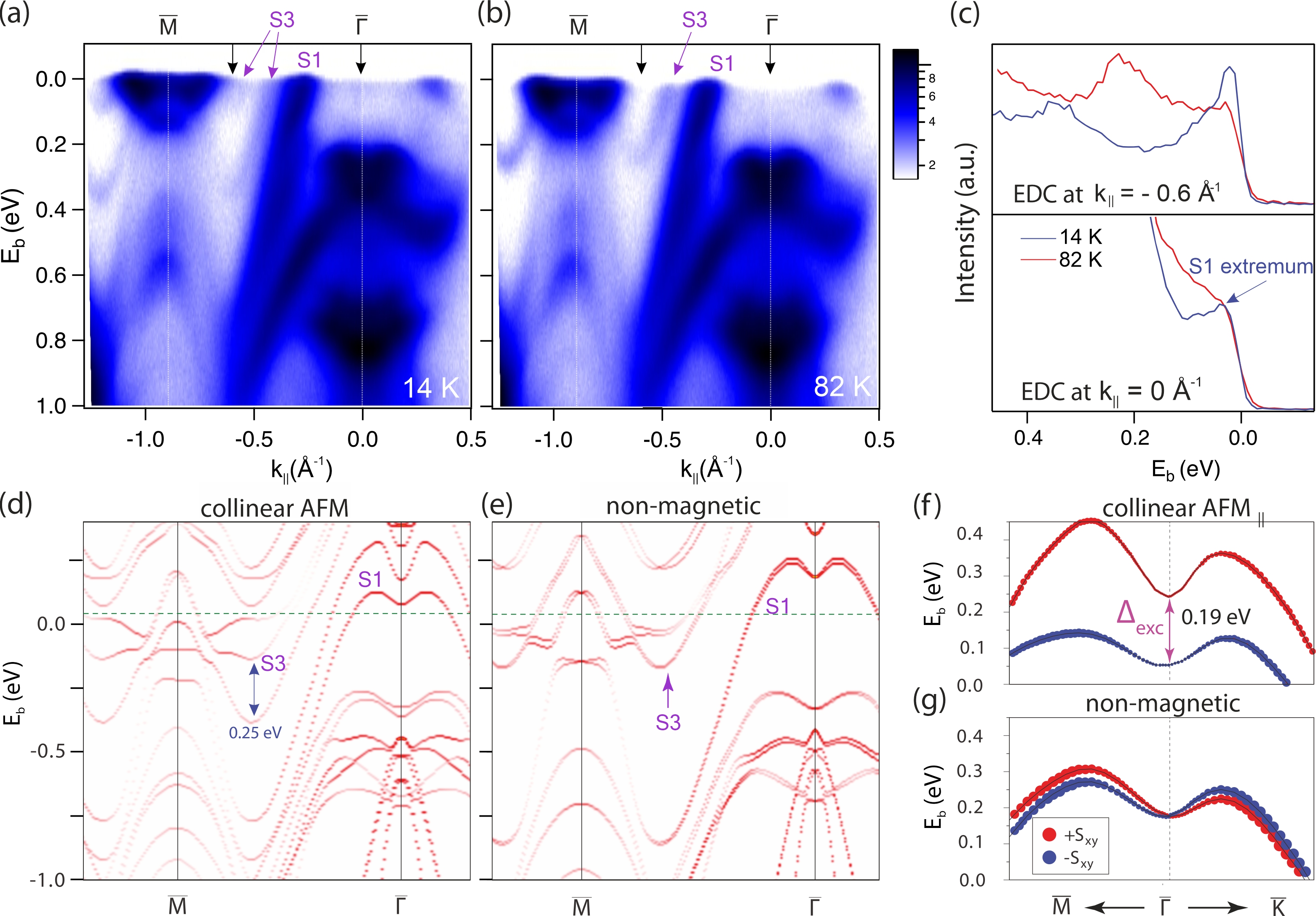}
\caption{Electronic structure of the 3 TL GdGe$_2$ film. (a, b) ARPES spectra recorded at 14 K and 82 K, respectively. A logarithmic intensity scale is used to highlight weak features.  (c) Energy distribution curves taken at k$_{||}$=0.6 \AA$^{-1}$ and $\bar{\Gamma}$ from the ARPES data of panel (a, b). (d, e) DFT calculations for the 3 TL  GdGe$_2$  film for cl  AFM$_\|$ and non-magnetic phases, respectively. The bands are unfolded onto the 1$\times$1 surface BZ to ease the comparison with the ARPES spectra. (f, g) Zoom of panels (d, e) near the top of the S1 band. Red and blue colors represent the opposite in-plane spin components.}
\label{f:triple}
\end{figure*}
The band structure calculation for the cl AFM$_\|$ and non-magnetic phases are shown in Figs.~\ref{f:double}(e, g) and Figs.~\ref{f:double}(f, h), respectively, while the band structure for the ncl 120${}^\circ$ AFM$_\|$ phase is shown in Fig.~\ref{f:double}(i). It should be noted that the unfolding procedure \cite{PRB-Band-UP-2014,PRB-Band-UP-2015} was used for comparison of the experimental and calculated band structures,   since collinear and ncl 120${}^\circ$ AFM$_\|$ band structures calculations require different unit cells. Namely, the cl AFM$_\|$  and non-magnetic band structures initially calculated within the   $3\times3$ and $\sqrt3\times\sqrt3$ supercells, respectively, were then unfolded onto the 1$\times$1 surface BZ. DFT calculations for cl AFM$_\|$ configuration (Fig.~\ref{f:double}(e)) perfectly reproduce the low-temperature features of the experimental electron dispersion (Fig.~\ref{f:double}(a)). The magnitude of the calculated exchange splitting of the S1 band at the $\bar{\Gamma}$ point is equal to 110 meV (Fig.~\ref{f:double}(e, g)), which is very close to the  ARPES observation of $\Delta\approx$130 meV at 14 K. Similarly, the calculated splitting of the S1 band at the  $\bar{\text{M}}$ point is 130 meV (Fig.~\ref{f:double}(e)) which is also close to the ARPES-derived value ($\Delta\approx$100 meV). The energy splittings predicted by DFT for the S2 state and for the cone-like feature at $\bar{\text{M}}$ are difficult to resolve by ARPES due to low intensity and a significant overlap with the 3 TL-derived features. The calculation for the non-magnetic case (Gd $4f$ orbitals are treated as core states) reproduces perfectly the experimental observation at temperatures above the magnetic transition (Fig.~\ref{f:double}(f)). The S1 band demonstrates a Rashba-like spin splitting \cite{Nat-Rev-Phys-Rashba-2022} in the vicinity of the $\bar{\Gamma}$  point with vortical spin-texture (Fig.~\ref{f:double}(h)). This Rashba-like spin splitting can be observed in the second derivative of 82 K ARPES spectrum in Fig.~\ref{f:double}(c). Instead, the band structure for the ncl 120${}^\circ$ AFM$_\|$ magnetic configuration (Fig.~\ref{f:double}(i)) does not show any significant exchange splitting for the S1 band at the $\bar{\Gamma}$ point.  This supports the total energy calculation results that ncl 120${}^\circ$ AFM$_\|$ magnetic configuration is not a ground state.
\par
Interestingly, the energy position of the exchange-split camel-like feature at $\bar{\Gamma}$ varies strongly with thickness. In the 3 TL GdGe$_2$ film it is located above the Fermi level (Figs.\ref{f:triple}(a,d) and the S1 band is not fully occupied, in contrast to the 2 TL GdGe$_2$ film. The magnetic transition here can still  be traced experimentally by analyzing the band structure evolution with temperature. The S3 band forms an electron pocket between $\bar{\Gamma}$ and $\bar{\text{M}}$ points with a minimum at k$_{||}$=0.6 \AA$^{-1}$ and exhibit exchange splitting: in the 14 K data there are two pockets separated by 270 meV with the lower branch minimum at ~340 meV  (Fig.~\ref{f:triple} (a)), while in the 82 K data there is single pocket with the minimum at ~250 meV (Fig.\ref{f:triple}(b)). The S3 band behavior in k$_{||}$=0.6 \AA$^{-1}$ can be better visualized in the top panel of Fig. \ref{f:triple}(c).  Another difference between the low and high temperature spectra is the appearance of a faint intensity at the Fermi level near the $\bar{\Gamma}$ point at 14 K that can be connected with the dip of the S1 band (bottom panel of Fig.\ref{f:triple}(c)). The behavior of these features at $\bar{\Gamma}$ and k$_{||}$=0.6 \AA$^{-1}$ is well reproduced by the calculations for the cl AFM$_\|$  and non-magnetic phases (Fig.~\ref{f:triple}(d,e)), in close analogy to the case of the 2 TL films. Also for the 3 TL film the quenching of the exchange splitting of the S1 band near $\bar{\Gamma}$  is accompanied by the appearance of Rashba-split bands with vortical spin-texture (Fig.~\ref{f:triple}(f,g)). \par
Notably, we did not observe explicit fingerprints of magnetic transition in the valence band structure of ultrathin DyGe$_2$ and GdSi$_2$ metalloxene films, which have the same crystallography structure of the GdGe$_2$ films (Fig. S3). DyGe$_2$ and GdGe$_2$ show nearly identical band structures. Consistently with the fact that bulk Dy$_3$Ge$_5$ has T$_\text{N}$=7 K, our measurements on the 2 TL DyGe$_2$ film at 14 K do not reveal any magnetically-induced band splitting. The band structure of the 2 TL GdSi$_2$ film presents the same features of the corresponding GdGe$_2$ film, with the only notable difference being the parabolic top of the S1 band at the $\bar{\Gamma}$ point. Despite bulk Gd$_3$Si$_5$ has T$_\text{N}$=54 K \cite{Roger2006}, which is higher than the value for bulk Gd$_3$Ge$_5$ (T$_\text{N}$=38 K), the ARPES data for the 2 TL GdSi$_2$ film do not display any signature of magnetically-induced band splitting. This behavior can be ascribed to the non-collinear AFM ground state of Gd$_3$Si$_5$ \cite{Roger2006}, which is likely to have only minor influence on the band structure, as  shown for the ncl 120${}^\circ$ AFM$_\|$ configuration of the 2 TL GdGe$_2$ film (Fig.~\ref{f:double}(i)). \par
\subsection{Uncompensated magnetic moments in  AFM GdGe$_2$  films}
\begin{table}[htbp]
\begin{center}
\begin{tabular}{|l|c|c|c|c|}
\hline 
   Atoms  & p- &  d- & f- & Total \\
\hline
   Ge-BL   &   0.045   & 0.000   & 0.000  & 0.045   \\
   Gd 1-st layer     &   0.010   & -0.136  & -6.917  & -7.042  \\ 
   Ge-interlayer  &   0.010   & 0.000   & 0.000  & 0.010   \\
   Gd 2-nd layer     &  -0.010   & 0.159   & 6.919  & 7.072   \\
   Ge(111)-BL     &   0.020   & 0.000   & 0.000  & 0.020   \\  
\hline
\end{tabular}
\end{center}
\caption{Orbital-decomposed Gd and Ge spin magnetic moment $S$ ($\mu_{\rm B}$) estimated for 2 TL GdGe$_{2}$ film with cl AFM$_\|$ magnetic ordering.}
\label{tab2}
\end{table}
As a final remark we want to address the discrepancy of our results with previously reported FM behavior in ultrathin metalloxene  films \cite{Tokmachev2019}. The 2 TL AFM GdGe$_2$ film can display an uncompensated magnetic moment induced by orbital hybridization of Gd and Ge atoms. Indeed, the S1 band that exhibits the largest splitting in the vicinity of the Fermi level in the AFM state is localized on both Gd and surrounding Ge atoms (Fig. 1S). As shown in Tab.~\ref{tab2} the expectation value of the spin moment ($S$) of the first Gd layer is -7.042 $\mu_{\rm B}$, while $S$ of the second Gd layer equals to 7.072 $\mu_{\rm B}$ resulting in the residual magnetic moment of 0.03 $\mu_{\rm B}$.  The Ge BL residing on top of the structure has $S$ = 0.045 $\mu_{\rm B}$; the $S$ of the upper Ge BL of the Ge(111) substrate is twice smaller, while  $S$ of the intermediate Ge BL is four times smaller. Hence, the Ge-related $S$ contribution is provided by the $p$ orbitals and is always positive. As a result the total $S$ of the  2 TL GdGe$_{2}$ is 0.105 $\mu_{\rm B}$ in spite of the AFM ground state ordering. This tiny but non-zero value is in line with the small magnetic moment (compared to that of Gd)  reported in the magnetic measurements for 2 TL thick metalloxene films \cite{Tokmachev2019}. Another possible explanation for the reported FM behavior is the unavoidable inhomogeneity of the  sample thickness that we observed directly in our samples and that was previously noticed in the scanning tunneling microscopy analysis of metalloxene films \cite{Wanke2009SS, ENGELHARDT2006755}. Space averaging investigation techniques, such as superconducting quantum interference device,  can pick up signals from different film thickness, including the FM 1 TL, and attribute the FM ground state to films of larger (nominal) thickness. \par
\section{Conclusions}	
Our study  of ultrathin rare-earth GdGe$_2$ metalloxenes films  by ARPES measurements and comprehensive \textit{ab initio} calculations shows in-plane FM order inside the single TL and AFM interlayer coupling. The hybridization between Gd and Ge orbitals induces a small uncompensated magnetic  moment in the AFM-ordered films, which was previously associated with a FM ground state for these systems.  The observed evolution of the GdGe$_2$ band structure with thickness suggests that exchange-split bands can be tuned in a broad energy range, thus providing a viable way to design the magnetic properties of the films, also through the well-known doping techniques of semiconductor technology. Another interesting way to tune the magnetism in such systems is the incorporation of the metalloxene magnetic layers within semiconductor  hetero-structures \cite{Bondarenko2021}, which makes them promising candidates for 2D materials engineering, similar to that proposed for layered chalcogenides. Finally, the spin \textit{S}, orbital \textit{L} and total magnetic momenta \textit{J} of the metalloxenes can be controlled by suitable choice of rare-earth elements that have similar valence states and ionic radius. All these aspects make metalloxens a convenient playground to study magnetism in the 2D limit.  

\section*{Author Contributions}
 A. V. Matetskiy: Investigation, Conceptualization, Supervision, Writing - original draft, Writing - Review and Editing, Project administration.
 V. Milotti: Investigation, Formal analysis, Writing - Review and Editing.
 P. M. Sheverdyaeva: Investigation, Validation, Writing - Review and Editing.
 P. Moras: Investigation, Validation, Writing - Review and Editing. 
 C. Carbone: Validation, Conceptualization.
 A. N. Mihalyuk: Software, Conceptualization, Visualization, Writing - original draft, Writing - Review and Editing. 

\section*{Conflicts of interest}
There are no conflicts to declare.

\section*{Acknowledgements}
A. N. Mihalyuk acknowledge that DFT calculations were supported by the Russian Science Foundation (Grant No. 19-12-00101, \url{https://rscf.ru/project/19-12-00101/}). A. V. Matetskiy, P. M. Sheverdyaeva, P. Moras,  C. Carbone acknowledge EUROFEL-ROADMAP ESFRI of the Italian Ministry of Education, University, and Research. The authors thank S.V. Eremeev for comments and helpful discussions. The calculations were conducted using the equipment of Shared Resource Center "Far Eastern Computing Resource" IACP FEB RAS (https://cc.dvo.ru).

\providecommand*{\mcitethebibliography}{\thebibliography}
\csname @ifundefined\endcsname{endmcitethebibliography}
{\let\endmcitethebibliography\endthebibliography}{}

\bibliographystyle{rsc}
\end{document}